\documentclass[pre,preprint,aps,eqsecnum]{revtex4}

\usepackage{graphicx}

\begin{document}

\title{Dynamics of Shear-Transformation Zones in Amorphous Plasticity: Formulation in Terms of an Effective Disorder Temperature}

\author{J. S. Langer}

\affiliation{Department of Physics,
University of California,
Santa Barbara, CA  93106-9530  USA}

\date{May, 2004}

\begin{abstract}
This investigation extends earlier studies of a shear-transformation-zone (STZ) theory of plastic deformation in amorphous solids.  My main purpose here is to explore the possibility that the configurational degrees of freedom of such systems fall out of thermodynamic equilibrium with the heat bath during persistent mechanical deformation, and that the resulting state of configurational disorder may be characterized by an effective temperature.  The further assumption that the population of STZ's equilibrates with the effective temperature allows the theory to be compared directly with experimentally measured properties of metallic glasses, including their calorimetric behavior. The coupling between the effective temperature and mechanical deformation suggests an explanation of shear-banding instabilities.
\end{abstract}

\maketitle

\section{Introduction}

This is the third in a recent series of studies of shear-transformation-zone (STZ) models of plastic deformation in amorphous solids.  In the earlier two papers, Falk, Pechenik, and I showed how to use principles of symmetry and energy balance to constrain the form of STZ theories at low temperatures \cite{LP}, and then used those ideas to construct a finite-temperature theory \cite{FLP} whose predictions could be compared with the behavior of metallic glasses observed by Kato et al.\cite{KATO} and Lu et al.\cite{LU}. Our version of STZ theory was introduced originally in \cite{FL}.  It differs from the flow-defect theories of Cohen, Turnbull, Spaepen, Argon and others \cite{TURNBULL,SPAEPEN77,SPAEPEN81,ARGON83} primarily in that, instead of simply postulating an equation of motion for the STZ density, we included a rudimentary model for the irreversible, internal dynamics of these zones.  This augmented STZ theory exhibits an exchange of dynamic stability between jammed and flowing states at a stress that we identify as a yield stress.  Our main conclusion in \cite{FLP} was that the transition between linear Newtonian viscosity and nonlinear superplasticity  reported in \cite{KATO,LU} can be explained quantitatively as a transition between thermally activated creep at low stress and the onset of plastic flow at the STZ yield stress. 

My first purpose here is to address several issues that were left outstanding in \cite{FLP}. The most important of these were questions pertaining to the STZ density.  In order to retain an essential simplicity in \cite{FLP}, we found it best to leave the theory in a form in which that density approached a temperature-independent value in the limit of small but nonvanishing driving force.  We pointed out that this limiting density, after indefinitely long aging, ought to relax to its thermal equilibrium value, and that we would need to incorporate such an aging mechanism into a next version of the theory.  More generally, we argued that the subtle interplay between limiting behaviors at small strain rates and small temperatures provides an important clue about the fundamental nonequilibrium properties of these systems.

My present hypothesis is that the STZ density is governed by an effective temperature, $T_{eff}$, of the kind discussed, for example, in papers by Ono et al. \cite{ONO}, Cugliandolo et al. \cite{CUGLIANDOLOetal}, Sollich et al. \cite{SOLLICHetal}, Berthier and Barrat \cite{BERTHIER-BARRAT}, and Lacks \cite{LACKS}.  Some aspects of these ideas are related to work by Mehta and Edwards \cite{ME}.  As proposed in Refs. \cite{ONO,CUGLIANDOLOetal,SOLLICHetal,BERTHIER-BARRAT,LACKS}, $T_{eff}$ differs from the ordinary thermal temperature $T$ in circumstances where the slowly changing configurational degrees of freedom of the system fall out of equilibrium with the heat bath -- a situation that occurs when molecular rearrangements are driven by plastic deformation. My ideas have emerged in part from discussions with Anael Lemaitre \cite{LEMAITRE}, who has approached the concept of effective temperature from a different point of view.  

More specifically, I assume that, in a nonequilibrium system, the STZ density is driven toward $n_{\infty}\,\exp\,(-1/\chi)$, where $\chi=k_B\,T_{eff}/E_Z$.  In this regard, the reduced effective temperature $\chi$ is very nearly, but not quite, the same as Spaepen's reduced free volume. \cite{SPAEPEN77,SPAEPEN81} Here, $E_Z$ is a characteristic energy associated with STZ formation, and $n_{\infty}$ is a density of the order of the number of molecules per unit volume.  This assumption implies that the local energy (or density) fluctuations of the slowly varying configurational degrees of freedom are described by a Boltzmann distribution with effective temperature $T_{eff}$. That is, I assume that persistent deformation accompanied by molecular rearrangements produces a steady state of disorder in an amorphous system.  In the absence of constraints other than number and energy conservation, the statistical distribution of density fluctuations must maximize an entropy, and therefore that distribution should be described by a temperature. To be consistent with the underlying assumption that the STZ's are sparsely distributed in the material, they must account for only a small fraction of the configurational degrees of freedom.  Thus the probability of finding STZ's should be accurately proportional to a Boltzmann factor, $\exp\,(-E_Z/k_B\,T_{eff})$, and should be far out in the wings of this statistical distribution.

The underlying structure of an STZ theory based on the effective-temperature hypothesis is outlined in Section II.  Here I reintroduce the fully nonlinear STZ theory \cite{FL} in a version suitable for use at nonzero temperatures.  As we pointed out in \cite{FLP}, the so-called ``quasilinear theory'' used in the preceding papers has serious shortcomings, specifically, lack of realistic memory effects and unrealistically large plastic deformation at small stresses.  It also has an unattractive asymmetry between the rates of shear transformations and dilational rearrangements, which must be corrected in order to deal systematically with thermal effects.  Section II concludes with a statement of the fully nonlinear equations of motion for the internal state variables. In Section III, I present arguments in favor of the effective-temperature hypothesis and make a rough estimate of the limiting value of $\chi$ at small but nonvanishing strain rates.  I then show how these arguments can be used to determine an equation of motion for  $\chi$.  

The remainder of this paper is devoted to exploring the predictions of these equations of motion and comparing them with the results of metallic glass experiments, primarily those of Lu et al. \cite{LU} on bulk amorphous ${\rm Zr_{41.2}\,Ti_{13.8}\,Cu_{12.5}\,Ni_{10}\,Be_{22.5}}$. I discuss a wide range of such measurements including the steady-state stress versus strain-rate data and the stress-strain curves obtained at various strain rates and temperatures, in analogy to our presentation in \cite{FLP}.  With this version of the theory, I can go on to compute specific heat curves obtained by differential scanning calorimetry, and also can discuss the way in which those measurements may be interpreted in terms of effective temperatures.  I conclude with some remarks about how shear banding instabilities may arise in theories of the kind introduced here.    

\section{Basic Equations of Motion}

Let us start by summarizing briefly the assumptions and definitions used in \cite{LP,FLP}.  Assume that, instead of being structureless objects as in the flow-defect theories of \cite{TURNBULL,SPAEPEN77,SPAEPEN81,ARGON83}, the STZ's are two-state systems which, in the presence of a shear stress, can transform back and forth between just two different orientations. Importantly, these STZ's are created and annihilated during irreversible deformations of the material.  As in \cite{FLP}, consider first a two-dimensional system, and subject it only to pure shear deformations. (The transformation of the two-dimensional results into a form suitable for analysis of three-dimensional experiments is discussed at the beginning of Section IV. It is described in more detail in \cite{FLP}.)  In this case, we need to consider only situations in which the orientation of the principal axes of the  stress and strain tensors remains fixed. That is, we do not need to consider situations in which a fully off-diagonal tensorial version of the STZ theory is necessary, as in the necking calculations reported in \cite{ELP}.  Therefore, it is sufficient to assume that the population of STZ's consists simply of zones oriented along the two relevant principal axes of the stress tensor and, without loss of generality, to let the deviatoric stress $s_{ij}$ be diagonal along the $x$, $y$ axes. Specifically, let $s_{xx}=-s_{yy}= s$ and $s_{xy}=0$. Then choose the ``$+$" zones to be oriented (elongated) along the $x$ axis, and the ``$-$" zones along the $y$ axis; and denote the population density of zones oriented in the ``$+$/$-$'' directions by the symbol $n_{\pm}$.  

With these conventions, the plastic strain rate is:
\begin{equation}
\label{epsdot}
\dot\epsilon_{xx}^{pl}= -\dot\epsilon_{yy}^{pl}\equiv \dot\epsilon^{pl}={\lambda\over\tau_0}\,\Bigl(R(-\tilde s)\,n_--R(\tilde s)\,n_+\Bigr).
\end{equation}
Here, $\lambda$ is a material-specific parameter with the dimensions of volume (or area in strictly two-dimensional models), which must have roughly the same order of magnitude as the volume of an STZ, that is, a few  cubic or square atomic spacings.  The remaining factor on the right-hand side of Eq.(\ref{epsdot}) is the net rate per unit volume at which STZ's transform from ``$-$'' to ``$+$'' orientations.  $R(\tilde s)/\tau_0$ and $R(-\tilde s)/\tau_0$ are the rates for ``$+$" to ``$-$'' and ``$-$" to ``$+$'' transitions respectively. The dimensionless stress is $\tilde s=s/\bar\mu$, where $\bar\mu$ is an effective shear modulus that will turn out to be an accurate approximation for the yield stress at the temperatures of interest here.  $\tau_0$ sets a time scale for the molecular rearrangements.  As we shall see, $\tau_0$ is not defined here in quite the same way as it was in \cite{FLP}. 

A basic assumption in this paper is that, in contrast to Eq.(3.3) in \cite{FLP}, we can rewrite the master equation for the densities $n_{\pm}$ in the form:
\begin{equation}
\label{ndot1}
\tau_0\,\dot n_{\pm}=R(\mp \tilde s)\,n_{\mp}-R(\pm \tilde s)\,\,n_{\pm}+  (\Gamma+\rho)\,\left({n_{\infty}\over 2}\,e^{-1/\chi}-n_{\pm}\right).
\end{equation}
The first pair of terms on the right-hand side describes the same switching back and forth of the STZ's that appears in Eq.(\ref{epsdot}), and the last terms describe the rates of creation and annihilation of zones.  In writing the latter terms, I am using the principle of detailed balance to fix the ratio of the annihilation and creation factors, and accordingly am omitting the quadratic term that we used in \cite{FLP}.  As before, the rate factor multiplying the annihilation and creation terms consists of the driven part $\Gamma$ and the spontaneous thermal part $\rho(T)$. 
Our usual notation is:
\begin{equation}
\label{vardef}
\Lambda \equiv {n_++n_-\over n_{\infty}},~~~~\Delta\equiv {n_+-n_-\over n_{\infty}};
\end{equation}
and
\begin{equation}
\label{Tdef}
{\cal S}(\tilde s)\equiv {1\over 2}\,\Bigl(R(-\tilde s)-R(+\tilde s)\Bigr),~~~~{\cal C}(\tilde s)\equiv {1\over 2}\,\Bigl(R(-\tilde s)+R(+\tilde s)\Bigr),~~~{\cal T}(\tilde s)\equiv {{\cal S}(\tilde s)\over{\cal C}(\tilde s)}.
\end{equation}
Then, using Eq.(\ref{epsdot}), and defining $\epsilon_0 \equiv \lambda\,n_{\infty}$, we have 
\begin{equation}
\label{doteps}
\tau_0\,\dot\epsilon^{pl}=\epsilon_0\,{\cal C}(\tilde s)\,\bigl(\Lambda\,{\cal T}(\tilde s)-\Delta\bigr);
\end{equation}
\begin{equation}
\label{dotdelta}
\tau_0\,\dot\Delta=2\,{\cal C}(\tilde s)\,\Bigl(\Lambda\,{\cal T}(\tilde s)-\Delta\Bigr)-(\Gamma +\rho)\,\Delta;
\end{equation}
and 
\begin{equation}
\label{dotlambda}
\tau_0\,\dot\Lambda=(\Gamma+\rho)\,\Bigl(e^{-1/\chi}-\Lambda \Bigr).
\end{equation}

The next step is to use the energy-balance argument introduced in \cite{LP} to evaluate the  quantity $\Gamma$.  Both the effective temperature and the fully nonlinear rate factor $R(\tilde s)$ will introduce features that were not present in \cite{LP} or \cite{FLP}; therefore it will be useful to rewrite this analysis.  As in the earlier papers, we start by writing the first law of thermodynamics in the form:
\begin{equation}
\label{energybalance}
2\,\dot\epsilon^{pl}\,s=
{2\, \epsilon_0\,\bar\mu\over\tau_0}\,{\cal C}(\tilde s)\,\Bigl(\Lambda\,{\cal T}(\tilde s)-\Delta\Bigr)\,\tilde s= {d\over dt}\,\Psi(\Lambda,\Delta) + {\cal Q}(\tilde s,\Lambda,\Delta).
\end{equation}
The left-hand side of Eq.(\ref{energybalance}) is the rate at which plastic work is done by the applied stress $s=\bar\mu\,\tilde s$.  On the right-hand side, $\Psi$ is the recoverable, state-dependent, internal energy associated with the STZ's.  Because the STZ's in this formulation represent only a very small fraction of the configurational degrees of freedom, $\Psi$ is not the energy (or enthalpy) obtained by calorimetric measurements.  Therefore, this picture is different from the one presented in \cite{LP,FLP}, where we did compare $\Psi$ qualitatively to calorimetric data.\cite{DEHEY,HB93,HB95}  In either case, $\Psi$ must be proportional to the density of STZ's, and must have the dimensions of energy per unit volume; therefore it is convenient to write it in the form:  
\begin{equation}
\Psi(\Lambda,\Delta)=\bar\mu\,\epsilon_0\,\Lambda\,\psi(m), ~~~~m\equiv{\Delta\over\Lambda}. 
\end{equation}

The last term on the right-hand side of Eq.(\ref{energybalance}), i.e. ${\cal Q}$, is the energy dissipation rate per unit volume.  The central hypothesis of \cite{LP} is that $\Gamma$ is simply proportional to the rate of energy dissipation per STZ.  That is,
\begin{equation}
\label{QGamma}
{\cal Q}(\tilde s,\Lambda,\Delta)= {\epsilon_0\,\bar\mu\over\tau_0}\,\Lambda\,\Gamma(\tilde s,\Lambda,m).
\end{equation}
We then can use Eqs.(\ref{dotdelta}) and (\ref{dotlambda}) to write Eq.(\ref{energybalance}) in the form:
\begin{eqnarray}
&&2\,{\cal C}(\tilde s)\,\Lambda\,\Bigl({\cal T}(\tilde s)- m \Bigr)\,\tilde s = \Bigl(\psi(m)-m\,\psi'(m)\Bigr)\,\Bigl(\Gamma(\tilde s,\Lambda, m)+\rho(T)\Bigr)\,\Bigl(e^{-1/\chi}-\Lambda\Bigr)\cr &&+ \,\,\psi'(m)\,\Lambda\,\Bigl[2\,{\cal C}(\tilde s)\,\Bigl({\cal T}(\tilde s)-m \Bigr)-\Bigl(\Gamma(\tilde s,\Lambda, m)+\rho(T)\Bigr)\,m \Bigr]+\Lambda\,\Gamma(\tilde s,\Lambda, m),
\end{eqnarray}
which can be solved easily for $\Gamma$ or, more conveniently, for $\tilde\Gamma\equiv\Gamma +\rho$:
\begin{equation}
\label{gammatilde}
\Gamma(\tilde s,\Lambda, m)+\rho(T)=\Lambda\,\left[{2\,{\cal C}(\tilde s)\,({\cal T}(\tilde s)-m)\,\Bigl(\tilde s-\psi'(m)\Bigr)+\rho(T)\over \Lambda-m\,\psi'(m)\,e^{-1/\chi}+\psi(m)\,\Bigl(e^{-1/\chi}-\Lambda\Bigr)}\right]\equiv \tilde\Gamma(\tilde s,\Lambda, m).
\end{equation}
Our equations of motion are now:
\begin{equation}
\label{doteps2}
\tau_0\,\dot\epsilon^{pl}=\epsilon_0\,{\cal C}(\tilde s)\,\Lambda\,\Bigl({\cal T}(\tilde s)-m\Bigr);
\end{equation}
\begin{equation}
\label{dotm2}
\tau_0\,\dot m = 2\,{\cal C}(\tilde s)\,\Bigl({\cal T}(\tilde s)-m\Bigr)-{m\over\Lambda}\,\tilde\Gamma(\tilde s,\Lambda, m)\,e^{-1/\chi};
\end{equation}
and
\begin{equation}
\label{dotlambda2}
\tau_0\,\dot\Lambda= \tilde\Gamma(\tilde s,\Lambda, m)\,\Bigl(e^{-1/\chi}-\Lambda\Bigr);
\end{equation}

Next we must specify the rate factors $R(\tilde s)$ and $\rho(T)$.  In \cite{FLP}, we wrote the dilational rate factor in the form
\begin{equation}
\label{rhoT}
{\rho(T)\over\tau_0}={\rho_0\over\tau_0}\,\exp\left(-{\Delta V_0^{dil}\over v_f(T)}\right),
\end{equation}
where $\rho_0$ is a dimensionless prefactor, $\Delta V_0^{dil}$ is the activation volume required for a dilational rearrangement, and $v_f(T)$ is usually identified as the free volume. In fact, in \cite{FLP} we treated $v_f(T)$ as a phenomenological function, not necessarily the same as the free volume, and evaluated it directly from the measured viscosities with no use of the Vogel-Fulcher or Cohen-Grest formulas.\cite{COHEN-GREST}  We then suggested, in analogy to the fully nonlinear STZ model \cite{FL}, that the shear rates ought to have the form
\begin{equation}
\label{Rnonlinear}
R(\tilde s) = \exp\left(-{\Delta V^{shear}(\tilde s)\over v_f(T)}\,\right);~~~~\Delta V^{shear}(\tilde s)= \Delta V_0^{shear}\,e^{-\tilde s},
\end{equation}
Setting the exponential prefactor equal to unity in Eq.(\ref{Rnonlinear}) defines $\tau_0^{-1}$ to be the value of the dimensional rate $R(\tilde s)/\tau_0$ in the limit $\tilde s\to\infty$.  Since the dominant temperature dependence of this rate should occur via the function $v_f(T)$ in the exponent, we may expect that $\tau_0$ is at most a slowly varying function of $T$.  With the definition $\Delta V_0^{shear}/v_f(T)\equiv \alpha(T)$, we have
\begin{equation}
\label{Cdef2}
{\cal C}(\tilde s)=\exp\left[-\alpha\,\cosh(\tilde s)\right]\,\, \cosh\left[\alpha\,\sinh(\tilde s)\right];
\end{equation}
and
\begin{equation}
\label{Tdef2}
{\cal T}(\tilde s) =\tanh\left[\alpha\,\sinh(\tilde s)\right].
\end{equation}
In the applications to be considered here, there seems to be no reason to expect two different activation volumes or two different time constants for dilational and shear rearrangements; therefore I shall assume that $\Delta V_0^{dil}=\Delta V_0^{shear}$ and $\rho_0=1$.  Then
\begin{equation}
\label{rho-alpha}
\rho(T)=e^{-\alpha(T)}.
\end{equation}
Eventually, we shall need to include pressure dependence in the function $\alpha(T)$; but that too will be unnecessary for present purposes.

The final step in deriving equations of motion for $\Lambda$ and $m$ is to choose $\psi(m)$ so that the numerator in the expression for $\tilde\Gamma$ in Eq.(\ref{gammatilde}) is non-negative for all values of $\tilde s$.  To do this, compute the inverse function of ${\cal T}$; that is, find the function $\xi(m)$ such that ${\cal T}(\xi)=m$.  Then, because ${\cal T}$ is a monotonically increasing function of its argument, the choice $\psi'(m)=\xi(m)$ assures us that both ${\cal T}(\tilde s)-m$ and $\tilde s-\xi(m)$ change sign at the same value of $\tilde s$, and therefore that the product of these two factors is never negative.  For the specific choice of ${\cal T}$ given in Eq.(\ref{Tdef2}), 
\begin{equation}
\label{sigma}
\xi(m)=\ln\,\left[\sqrt{1+{1\over 4\,\alpha^2}\,\ln^2 \Bigl({1+m\over 1-m}\Bigr)}+{1\over 2\,\alpha}\,\ln\, \Bigl({1+m\over 1-m}\Bigr)\right];
\end{equation}
and
\begin{equation}
\psi(m)=\psi(0)+\int_0^m\,\xi(m)\,dm.
\end{equation}
$\psi(0)$ is an as-yet undetermined constant which, as it turns out, we shall not need to evaluate. The result is
\begin{equation}
\tilde\Gamma(\tilde s,\Lambda,m)={\Lambda\over M(\Lambda,m)}\,\left[2\,{\cal C}(\tilde s)\,({\cal T}(\tilde s)-m)\,\Bigl(\tilde s-\xi(m)\Bigr)+\rho(T)\right],
\end{equation}
where
\begin{equation}
M(\Lambda,m)= \Lambda-m\,\xi(m)\,e^{-1/\chi}+\psi(m)\,\Bigl(e^{-1/\chi}-\Lambda\Bigr).
\end{equation}

Positivity of $\tilde\Gamma$ requires that $M(\Lambda,m)$ remain positive along all the system trajectories determined by Eqs.(\ref{dotm2}) and (\ref{dotlambda2}) in the space of variables $\Lambda$ and $m$. This happens automatically so long as all the trajectories start at points where $M(\Lambda,m)>0$.  The locus of points along which $M(\Lambda,m)$ changes sign is a dynamical boundary for these trajectories; the dissipation rate diverges at that boundary, and the trajectories are strongly repelled from it in a way that does not allow them to cross into unphysical regions where the dissipation rate is negative.  An interesting feature of this fully nonlinear case is that the boundary always occurs when $m$ is slightly smaller than unity because of the weak divergence of the function $\xi(m)$ when $m\to 1$.  For example, setting $\Lambda = e^{-1/\chi}$ and using Eq.(\ref{sigma}) with $\alpha = 2$, I find that the upper limit of $m$ is 0.983732.  We shall see that the interesting values of $\alpha$ are generally much larger than this, of order 10 or more, in which case the upper limit of $m$ is practically indistinguishable from unity.

These equations of motion simplify greatly if we note that $\Lambda = e^{-1/\chi}$ is always the only stable stationary solution of Eq.(\ref{dotlambda2}), and use this relation to eliminate $\Lambda$ from the beginning of the analysis.  (There seems to be no conventional experimental method for adjusting $\Lambda$ and $\chi$ independently of one another.) Then we have
\begin{equation}
\label{doteps3}
\tau_0\,\dot\epsilon^{pl}=\epsilon_0\,e^{-1/\chi}\,{\cal C}(\tilde s)\,\Bigl({\cal T}(\tilde s)-m\Bigr);
\end{equation}
\begin{equation}
\label{dotm3}
\tau_0\,\dot m = {2\,{\cal C}(\tilde s)\,\Bigl({\cal T}(\tilde s)-m\Bigr)\,(1-m\,\tilde s)-m\,\rho(T)\over 1-m\,\xi(m)};
\end{equation}
and
\begin{equation}
\label{gammatilde3}
\tilde\Gamma={2\,{\cal C}(\tilde s)\,\Bigl({\cal T}(\tilde s)-m\Bigr)\,\Bigl(\tilde s-\xi(m)\Bigr)+\rho(T)\over 1-m\,\xi(m)}.
\end{equation}
Note that the effective temperature $\chi$ now appears explicitly only in the strain-rate equation, Eq.(\ref{doteps3}), where it determines the density of STZ's that must appear in front of the rate factor.  Conveniently, the as-yet undetermined energy $\psi(0)$ disappears entirely when we assume that $\Lambda$ is always in equilibrium with the configurational degrees of freedom at their effective temperature.  Note also that this approximation will have no effect  on any of the steady-state calculations presented below. 

Equations (\ref{doteps3}) and (\ref{dotm3}) describe the same exhange of stability at a yield stress that we found in earlier papers \cite{FL,LP,FLP}.  At low temperatures, where $\rho(T)\to 0$, the steady-state solutions of Eq.(\ref{dotm3}) are $m={\cal T}(\tilde s)$ (the jammed state with $\dot\epsilon^{pl}=0$), and $m=1/\tilde s$ (the flowing state with $\dot\epsilon^{pl}\ne 0$).  The two curves cross at $\tilde s = \tilde s_y$ where $\tilde s_y\,{\cal T}(\tilde s_y)=1$.  The jammed state is dynamically stable for $\tilde s < \tilde s_y$, and the flowing state is stable for $\tilde s > \tilde s_y$. For values of $\alpha(T)$ appreciably larger than unity, the solution of this equation is $\tilde s_y\approx 1$; thus the yield stress is $s_y\approx \bar\mu$.  

For nonzero $\rho(T)$, the stable branch of the steady-state solutions of Eq.(\ref{dotm3}), say $m=m_0(\tilde s)$, is
\begin{eqnarray}
\label{mequation}
\nonumber
m_0(\tilde s)={1\over 2\,\tilde s}\,\Bigl(1&+& \tilde s\,{\cal T}(\tilde s)+{\rho(T)\over 2\,{\cal C}(\tilde s)}\Bigr)\cr\\ &-& {1\over 2\,\tilde s}\sqrt{\Bigl(1+ \tilde s\,{\cal T}(\tilde s)+{\rho(T)\over 2\,{\cal C}(\tilde s)}\Bigr)^2 - 4\,\tilde s\,{\cal T}(\tilde s)}\,.
\end{eqnarray}
This function is shown in Fig.(1) along with graphs of $m={\cal T}(\tilde s)$ and $m=1/\tilde s$, which are the asymptotic values of $m_0(\tilde s)$ in the limit $\rho\to 0$, below and above the yield stress respectively.  In order that these two sets of curves not lie exactly on top of each other, I have chosen $\alpha=6$, which will turn out to correspond to a relatively high temperature of about $730\,K$, and have used Eq.(\ref{rho-alpha}) to evaluate $\rho$ in Eq.(\ref{mequation}).  

\section{Effective Temperature}

We now need an equation of motion for the effective temperature $\chi$.  Before writing such an equation, however, it will be useful to discuss some underlying concepts.

The theory as described so far contains three distinct time scales.  The first of these is $\tau_0$, the roughly temperature-independent time associated with STZ transitions that are driven by stresses of order the yield stress or larger.  This time will turn out to be very short, of order microseconds.  A second time scale is $\tau_0/\rho(T)\equiv \tau_T$, which is the strongly temperature-dependent time associated with spontaneous (i.e. stress-independent) thermally activated molecular rearrangements.  At low temperatures, $\tau_T$ becomes very long.  The third time scale is the inverse of the strain rate, $(\dot\epsilon^{pl})^{-1}\equiv \tau_{eps}$, which is determined by the externally imposed loading.  

We argued in \cite{FLP} that $\tau_{eps}$ is the only relevant time scale for  behavior in the regime where $\tau_0 \ll \tau_{eps} \ll \tau_T$.  This is the situation in which the temperature $T$ is so low that molecular rearrangements are not thermally activated, and the actual rearrangements, when they occur, are effectively instantaneous.  Thus, under steady-state conditions, the number of events in which rearrangements occur is not proportional to the time but to the strain.  As already stated in the Introduction, steady-state deformation with molecular rearrangements must produce a steady state of disorder in an amorphous system -- a statistical distribution of density fluctuations which, in turn, ought to maximize an entropy and therefore be described by an effective temperature. Therefore, after very long times, and so long as $\tau_{eps} \ll \tau_T$, the quantity $\chi$ must approach a definite value, say $\chi_{\infty}$. A more mathematically precise way of saying this, which reminds us that we must be looking in the limit in which both $\tau_{eps}$ and $\tau_T$ are very much longer than $\tau_0$, is that $\chi\to\chi_{\infty}$ if we take the limit $\dot\epsilon^{pl}\to 0$ {\it after} $T \to 0$.

To make a rough estimate for $\chi_{\infty}$, we can use the Stokes-Einstein fluctuation-dissipation relation in a way that will require careful discussion.   An oversimplified derivation starts by noting that, because there is a yield stress $s_y$ in these models, the viscosity is $\eta =s_y/(2\,\dot\epsilon^{pl})$.  Then, if the only relevant time scale in the model is $\tau_{eps}$, it follows that the diffusion constant $D$, measured over times much longer than $\tau_{eps}$, must be proportional to $\ell^2\,\dot\epsilon^{pl}$, where $\ell$ is the characteristic displacement of a molecule during an STZ-like rearrangement, i.e., roughly a molecular spacing.  Finally, the Stokes-Einstein relation says that $D\propto k_B\,T_{\infty}/\eta\,\ell$, where $T_{\infty}= E_Z\,\chi_{\infty}/k_B$. It follows that $k_B\,T_{\infty} \propto \ell^3\,s_y$, independent of $\dot\epsilon^{pl}$. 

One problem with this analysis is that the viscosity in the Stokes-Einstein formula is the linear response coefficient relating flow to driving force in the limit of vanishing stress and strain rate, whereas it is used here at the yield stress.  A related and possibly mitigating problem is that the ``temperature'' $T_{\infty}$ that we supposedly are evaluating with the Stokes-Einstein formula is a very low-frequency (essentially static) noise strength that determines the spatial distribution of energy and density fluctuations but not the rates at which those fluctuations vary in time.  The temporal rates are determined by the strain rate $\dot\epsilon^{pl}$, which, in this limit, is small but much faster than the essentially negligable rearrangement rates induced by true thermal fluctuations.

It helps to visualize the situation as follows.  At zero $T$, for small strain rates, the graph of stress as a function of strain-rate consists simply of a horizontal line at $ s=s_y$.  Now add to this system a slow noise source with characteristic frequencies of order, say, $\omega_{noise}$, which couples only to the configurational degrees of freedom.  Let the strength of this noise source be determined by an effective temperature $T_{\infty}$.  At non-zero $T_{\infty}$, the stress versus strain-rate graph must start at the origin and rise with a slope $2\,\eta(T_{\infty})$, where $\eta(T_{\infty})$ is the viscosity measured by averaging the stress over times longer than $\omega_{noise}^{-1}$. This section of the curve at small $\dot\epsilon^{pl}$ will extrapolate to $s=s_y$ at a strain rate, say, $\dot\epsilon^{pl}(T_{\infty}) = s_y/(2\,\eta(T_{\infty}))$.  For strain rates $\dot\epsilon^{pl} > \dot\epsilon^{pl}(T_{\infty})$, the curve returns to $s=s_y$. The Stokes-Einstein relation pertains to the portion of this function that goes into the origin at small strain rates.  Thus, so long as we measure diffusion and viscosity over times longer than $\omega_{noise}^{-1}$, we can estimate the diffusion constant: $D(T_{\infty})\propto \ell^2\,\dot\epsilon^{pl}(T_{\infty})$ and conclude that $k_B\,T_{\infty}\propto \ell^3\,s_y$ as before, independent of what value of $\dot\epsilon^{pl}=\dot\epsilon^{pl}(T_{\infty})$ we chose at the beginning. 

We can use the above argument -- not much more than a dimensional analysis -- to make an order-of-magnitude estimate for $T_{\infty}$.  The tensile yield stress that we used in \cite{FLP} was 1.9 GPa, which gives us an approximate value for $s_y$.  Then $T_{\infty}\sim 50\,\ell_A^3\,K$, where $\ell_A$ is the molecular length scale $\ell$ measured in Angstroms. If $\ell_A \sim 3$, then $T_{\infty}\sim 10^3\,K$.  A similar rough estimate emerges if we guess on dimensional grounds that $E_Z\sim \mu\,\ell^3$, where $\mu$ is the shear modulus.  Then, using the value of Young's modulus given in \cite{LU}, we have $\chi_{\infty}\sim s_y/\mu \sim 0.02$.  If $E_Z \sim 2\,ev$, then again we find $T_{\infty}\sim 10^3\,K$.  These estimates are consistent with Andrea Liu's suggestion \cite{ALIUTg} that $T_{\infty}$ is the glass temperature, which also is roughly of the order of $10^3\,K$ for these materials. 

With this understanding of the role and approximate magnitude of the effective temperature, we now can deduce an equation of motion for it by returning to the principle of energy balance.  As noted above, one of the main differences between this model and that discussed in \cite{FLP} is that, here, the energy stored in the STZ's is only a very small fraction of the energy contained in the configurational degrees of freedom. Thus we can assume that the energy dissipated by the STZ's during plastic deformation simply adds to the energy of configurational disorder. It then seems reasonable to assume that, over the range of temperatures of interest here (approximately 550 K - 700 K for the data reported in \cite{LU}), the specific heat of the configurational degrees of freedom is a constant, say, $C_D=k_B\,c_0/\ell^3$, where $c_0$ is a dimensionless number of order unity.  The associated configurational energy is $C_D\,T_{eff}$; and the energy-balance equation, i.e. the equation for the rate at which this energy is changing per unit time, $C_D\,\dot T_{eff}$, becomes the equation of motion for $T_{eff}$. 

In accord with the discussion in the preceding paragraphs, I propose to write this equation of motion in the form:
\begin{equation}
\label{Tdot}
C_D\,\dot T_{eff}={\cal Q}\,\Bigl(1-{T_{eff}\over T_{\infty}}\Bigr)+\epsilon_0\,K(\chi)\,{\rho(T)\over\tau_0}\,k_B\,(T-T_{eff}).
\end{equation}
The first term on the right-hand side says that the energy dissipated during plastic deformation, at rate ${\cal Q}$ per unit volume, is absorbed by the configurational degrees of freedom.  The second term proportional to ${\cal Q}$ is the one that says that this process must drive the system toward a limiting state of disorder in which $T_{eff} \to T_{\infty}$.  This equation can be used only when time variations are very much slower than the microscopic rate $\tau_0^{-1}$, because the preceding argument for a limiting value of $T_{eff}$ is valid only in those circumstances. Thus, we cannot expect this theory to be accurate for strain rates higher than about $(\epsilon_0/\tau_0)\,\exp\,(-1/\chi_{\infty})$; however, such rates are well above the experimental range.  

The term in Eq.(\ref{Tdot}) proportional to $\rho(T)$ says that $T_{eff}\to T$ in the absence of external driving, and does so at a rate which becomes very small at low temperatures. $K(\chi)$ is an equilibration coefficient, defined with a factor $\epsilon_0$ for convenience. The $\chi$ dependence of $K$ reflects the fact that the equilibration rate must depend on the state of disorder.  In what follows, I shall assume that 
\begin{equation}
\label{Kdef}
K(\chi)=\kappa\,e^{-\beta/\chi},
\end{equation}
so that $\epsilon_0\,K(\chi)$ is proportional to the density of of sites at which the equilibration transitions take place.  It will be simplest at first to let the equilibration parameter $\beta=1$, which means that the latter sites are the same as the STZ's.  However, as we shall see, other possibilities are interesting. 

Next, convert Eq.(\ref{Tdot}) into an equation for $\chi$ by writing, as in Eq.(\ref{QGamma}),
\begin{equation}
{\cal Q}(s,\Lambda,\Delta)={\bar\mu\,\epsilon_0\over \tau_0}\,e^{-1/\chi}\,\Gamma(\tilde s,m).
\end{equation}
The function $\Gamma$ (not $\Gamma+\rho$) is
\begin{equation}
\label{gamma3}
\Gamma(\tilde s,m)={2\,{\cal C}(\tilde s)\,\Bigl({\cal T}(\tilde s)-m\Bigr)\,\Bigl(\tilde s-\xi(m)\Bigr)+\rho(T)\,m\,\xi(m)\over 1-m\,\xi(m)}.
\end{equation}
Note that the term proportional to $\rho(T)$ in $\Gamma$ disappears in an undriven system because $m\to 0$ in that case. We then find
\begin{equation}
\label{chidot}
{\tau_0\,c_0\over\epsilon_0}\,\dot\chi= e^{-1/\chi}\,\Gamma(\tilde s,m)\,(\chi_{\infty}-\chi) + \kappa\,\rho(T)\,e^{-\beta/\chi}\,\left({T\over T_Z}-\chi\right).
\end{equation}
In order to avoid adding another arbitrary constant of order unity, I have used $\chi_{\infty}=\bar\mu\,\ell^3/E_Z$ in evaluating the coefficient of $\Gamma$ in Eq.(\ref{chidot}). Equation (\ref{chidot}), along with Eqs.(\ref{doteps3}), (\ref{dotm3}), and (\ref{gamma3}), provides a complete specification of the equations of motion for this model. 

\section{Limiting Behaviors at Small Stress}

At this point in the development, it is necessary to rewrite the two-dimensional STZ equations of motion in a form in which they can be applied directly to  three-dimensional experiments, especially those reported in \cite{LU}.  To do this, I assume that the stresses and strain rates are uniform throughout the experimental samples, and follow \cite{FLP} and \cite{ELP} by assuming that I can simply replace the variables $\tilde s$, $\dot\epsilon^{pl}$, and $m$ by traceless symmetric tensors.  In the case of a uniform sample with uniaxial applied stress in, say, the $x$ direction, and free, stressless surfaces normal to the $y$ and $z$ axes, each of these tensors is diagonal with elements proportional to $(1,-1/2,-1/2)$.  The total stress tensor $\sigma_{ij}$ has only one nonzero element, $\sigma_{xx}\equiv \sigma$.  Define $m^2 = (1/2)\,m_{ij}\,m_{ij} = (3/4)\,m_{xx}^2$, so that $m=\sqrt{3/4}\,m_{xx}$; and, similarly, $\tilde s^2 =(1/2)\,\tilde s_{ij}\,\tilde s_{ij} = (3/4)\,\tilde s_{xx}^2$, $\tilde s = \sqrt{3/4}\,\tilde s_{xx}$.  The only way in which this analysis differs from that in \cite{FLP} is that, here, we must form tensorial generalizations of the functions ${\cal T}(\tilde s)$ and $\xi(m)$.  This can be done most simply by writing ${\cal T}_{ij}(\tilde s) = (\tilde s_{ij}/|\tilde s|)\,{\cal T}(\tilde s)$, and $\xi_{ij}(m) =(m_{ij}/|m|)\,\xi_{ij}(m)$.  

With these transformations, we recover precisely our earlier formulas, Eqs.(\ref{doteps3}), (\ref{dotm3}), (\ref{gamma3}), and (\ref{chidot}) as the $xx$ components of the tensor equations. The single difference is that, because the experimental strain rate $\dot\epsilon_{xx}^{pl}$ is not rescaled as are $\tilde s$ and $m$, the parameter $\epsilon_0$ in Eq.(\ref{doteps3}) is replaced by $\epsilon_0'=\sqrt{4/3}\,\epsilon_0$.  The low-temperature exchange of stability still occurs when $\tilde s = \tilde s_y$ where $\tilde s_y\,{\cal T}(\tilde s_y)=1$; therefore, at the temperatures of interest here, we still have $\tilde s_y \cong 1$, $s_y \cong \bar\mu$.  The experimental data is expressed in terms of the tensile stress, which becomes $\sigma = (3/2)\,s_{xx} = \sqrt{3}\,\bar\mu\,\tilde s = \sigma_y\,\tilde s$, where $\sigma_y$ is the tensile yield stress.  

The first quantity that we must compute  is the Newtonian viscosity $\eta_N$, that is, the linear viscosity in the limit of vanishingly small stress and strain rate.  As in \cite{FLP}, comparing our theoretical $\eta_N$ with the experimental measurements reported in \cite{LU} provides initial constraints on several of the parameters that appear in our equations. In the small-$\tilde s$ limit, we have ${\cal T}(\tilde s)\approx \alpha\,\tilde s$, and ${\cal C}(\tilde s)\approx {\cal C}(0)= \exp(-\alpha)$.  Then we can deduce from Eqs.(\ref{Tdef2}), (\ref{rho-alpha}) and (\ref{mequation}) that, to lowest (linear) order in $\tilde s$, 
\begin{equation}
m_0(\tilde s) \approx {2\over 3}\,\alpha\,\tilde s.
\end{equation}
Using  Eq.(\ref{dotm3}), and noting that the product $m_0(\tilde s)\,\tilde s$ is small of order $\tilde s^2$, we have  
\begin{equation}
2\,{\cal C}(\tilde s)\,\Bigl({\cal T}(\tilde s)-m_0(\tilde s)\Bigr)\approx m_0(\tilde s)\,\rho(T);
\end{equation}
so  that, using Eq.(\ref{doteps3}), we find
\begin{equation}
\dot\epsilon_{xx}^{pl}={\epsilon_0\over\tau_0}\,e^{-1/\chi}\,{\cal C}(\tilde s)\,\Bigl({\cal T}(\tilde s)-m_0(\tilde s)\Bigr)\approx {\epsilon_0'\over 2\,\tau_0}\,e^{-1/\chi}\,m_0(\tilde s)\,\rho(T).
\end{equation}
To evaluate $\chi$ from Eq.(\ref{chidot}), note from Eq.(\ref{gamma3}) that $\Gamma$ is small of order $\tilde s^2$, so that $\chi\approx T/T_Z$ in the small-$\tilde s$, steady-state limit. Therefore,
\begin{equation}
\dot\epsilon_{xx}^{pl}\approx {\epsilon_0'\over 3\,\tau_0}\,\alpha(T)\,\tilde s\,\exp\,\left[-{T_Z\over T}-\alpha(T)\right],
\end{equation}
and 
\begin{equation}
\label{etaN}
\eta_N(T)=\lim_{\dot\epsilon^{pl}\to 0}\,{\bar\mu \tilde s_{xx}\over 2\,\dot\epsilon_{xx}^{pl}}={\sqrt{3}\,\bar\mu\,\tau_0\over \epsilon_0'\,\alpha(T)}\,\exp\,\left[{T_Z\over T}+\alpha(T)\right].
\end{equation}

We also can use this analysis to compute the temperature-dependent stress relaxation rates discussed in \cite{LU}. In these measurements, samples  first were compressed at relatively small strain rates and then held at fixed total strain $\epsilon^{total}$ while the stress was measured as a function of time.  The equation of motion that we must solve therefore is
\begin{equation}
\dot\epsilon_{xx}^{total}= {\sigma_y\over E}\,\dot{\tilde s} +\dot\epsilon_{xx}^{pl} =0,
\end{equation}
where $E$ is Young's modulus.  Using the preceding small-stress approximations, and again assuming that $\chi$ is thermalized in these experiments ($\chi\approx T/T_Z$), we find
\begin{equation}
\label{s-relax}
\dot{\tilde s} \approx - {\epsilon_0'\,E\over 3\,\tau_0\,\sigma_y}\,\alpha(T)\,\exp\,\left[-{T_Z\over T}-\alpha(T)\right]\,\tilde s \equiv -{\tilde s\over t_r(T)},
\end{equation}
which implies that the exponential relaxation time is
\begin{equation}
t_r(T) \approx {\eta_N(T)\over E}.
\end{equation} 
This relation is consistent with the conclusion of Lu et al. \cite{LU} that both $\eta_N(T)$ and $t_r(T)$ scale with the same temperature dependent rate factor. However, if $E = 96$ GPa as reported by \cite{LU}, these theoretical values of $t_r$ are too small by a factor of about 50.  This discrepancy may be due to the thermalization assumption, which might not be consistent with the way in which these measurements were made.  
 
In \cite{FLP}, we used our expression for $\eta_N(T)$ and the experimental values for this quantity given in \cite{LU} to obtain estimates of $\rho(T)$ (up to a scale factor) for eight separate values of the temperature.  However, our present formula for the Newtonian viscosity, Eq.(\ref{etaN}), is  more complicated than the one in \cite{FLP} because it now contains a temperature-dependent STZ density, proportional to $\exp\,(-1/\chi)\approx \exp\,(-T_Z/T)$, as well as the temperature-dependent rate factor $\rho(T)$.  Moreover, it will be useful for present purposes to have a smooth functional representation of $\rho(T)$ rather than just values at separate points.  Accordingly, I have fit Eq.(\ref{etaN}) to the experimental data using the Cohen-Grest formula \cite{COHEN-GREST} as a purely phenomenological fitting function for $\alpha(T)$. That is, in Eq.(\ref{rho-alpha}), I have used
\begin{equation}
\label{CGalpha}
\alpha(T)={T_R\over T-T_0+\sqrt{(T-T_0)^2+T_1\,T}},
\end{equation}
where $T_R$, $T_0$, and $T_1$ are fitting parameters with the dimensions of temperature. 

Clearly, we cannot obtain a unique fit for all the parameters in Eqs.(\ref{etaN}) and (\ref{CGalpha}) from just the viscosity data, so we now must make some  physical assumptions.  The guiding principle is to make the simplest  possible choices, and to add complications only if they become necessary.  In this spirit, we may assume that there is only one temperature that characterizes the glass transition.  In Eq.(\ref{CGalpha}), that temperature is $T_0$.  If we then adopt Liu's hypothesis \cite{ALIUTg} that $T_{\infty}$ is the glass temperature,  we should choose $T_{\infty}=T_0$.  On the basis of various clues, including calorimetric analyses, I estimate that $T_{\infty}=T_0 \cong 800\, K$. This value is consistent with my guess that the temperatures used in the experiments of \cite{LU} are all well below $T_0$; that is, the behaviors seen in these experiments seem characteristic of states in which the material is softening rapidly with increasing $T$ but is still stiff enough to resemble a solid in resisting deformation. 

Next we must estimate $T_Z$.  In the preceding dimensional analysis, we guessed that $T_Z/T_{\infty}\sim \mu/s_y \sim 50$, which would imply that $T_Z\sim 40,000\,K$.  This estimate, however, is uncertain by a least a factor 2.  A better strategy, I think, is to assume that the Newtonian viscosity is dominated at the higher temperatures shown in \cite{LU}, Fig.10,  by the factor $\exp\,(T_Z/T)$ in Eq.(\ref{etaN}). That fitting strategy yields $T_Z \cong 25,000\,K$, which is within the previous uncertainty and is the value that I will use here.  It means that $E_Z\cong 2$ ev, which seems plausible for a vacancy-formation energy.

Figure (2) shows the fit to the Newtonian viscosity as a function of temperature with $T_Z=25,000\,K$, $T_0=800\,K$, $T_R=600\,K$, $T_1=28\,K$, and $\eta_0 \equiv \bar\mu\,\tau_0/ \epsilon_0 \cong 2 \times 10^{-11}\,{\rm Pa\,sec}$.  The eight points at the lowest temperatures in Fig.(2) are the ones that we used in \cite{FLP}; the four points at higher temperatures are also taken from \cite{LU}, Fig.(10).   Figure (3) shows the corresponding function $\alpha(T)$. The values of $\alpha$ in the range of experimental interest, roughly $600\,K$ to $700\,K$, are  of order $8$ to $15$, that is, in about the same range as the values that Falk and I \cite{FL} found to fit the original MD simuations.

As in \cite{FLP}, I assume that the tensile yield stress at the experimental temperatures is the same as the room temperature value reported in \cite{LU}, i.e $\sigma_y= $ 1.9 GPa.  Thus $\bar\mu\, \sigma_y/\sqrt{3} \cong$ 1.1 GPa.  With the above value of $\eta_0$, we have $\epsilon_0'/\tau_0 \cong 6.3\times 10^{19}\,{\rm sec}^{-1}$. 

\section{Analysis and Comparison with Experiments}

We are now ready to explore the properties and experimental predictions of this effective temperature theory at values of the stress and strain rate where the response to loading becomes nonlinear. Look first at the steady-state solutions obtained by using Eq. (\ref{doteps3}) to compute the strain rate (with $\epsilon_0\to \epsilon_0'$), and by setting the time derivatives on the left-hand sides of Eqs. (\ref{dotm3}) and (\ref{chidot}) to zero.  The stresses and strain rates found in this way correspond to those obtained by Lu et al. \cite{LU} from the late, steady-state stages of their constant-strain-rate measurements.  The steady-state values of the reduced effective temperature $\chi$ may, in principle, be obtained by calorimeteric measurements as discussed below.  

It is simplest to start by setting $\beta=1$ in Eq.(\ref{Kdef}) which, as mentioned earlier, means that the thermal fluctuations that drive the effective disorder temperature toward the temperature of the heat bath occur predominantly at the STZ sites.  The only other adjustable parameter in steady state is $\kappa$.  Figure (4) shows the dimensionless stress $\tilde s$ as a function of the scaled strain rate $\eta_N(T)\,\dot\epsilon_{xx}^{pl}$ for $\beta=1$, $\kappa = 2$, and for four different temperatures $T$ in the range of the metallic glass data \cite{LU}.  It should be compared with Fig.(3) of \cite{FLP}, in which these curves lie accurately on top of one another up through the yield stress.  This scaling behavior, which was discovered experimentally by Kato et al. \cite{KATO} and explored in more detail by Lu et al. \cite{LU}, clearly is broken here.  The trend toward lower stresses at lower temperatures can be understood as a nonlinear property of the effective temperature theory.  As the strain rate increases, $T_{eff}$ increases, and the driving force required to maintain that strain rate decreases accordingly.  Because $\eta_N(T)$ increases rapidly with decreasing $T$, we may understand the curves that are plotted as functions of $\eta_N(T)\,\dot\epsilon_{xx}^{pl}$ in Fig.(4) to be a sequence in which $T_{eff}$ increases as $T$ decreases. 

The important question is whether this non-scaling behavior is ruled out by experiment.  Figure (5) shows a direct comparison of the data from \cite{LU} with theoretical curves for eight different temperatures, in analogy to Fig.(5a) in \cite{FLP}.  The value $\kappa=2$ was chosen to optimize the fit to the data at $643\,K$.  If we use the temperatures cited in \cite{LU}, the agreement is reasonably accurate for $623\,K$ and above (apart from a few apparently outlying points), and also (perhaps fortuitously) is satisfactory for the two points at $573\,K$.  The theoretical curves for the latter set of temperatures are shown by solid lines in the figure.  

However, the agreement is not so good at $593\,K$ and $603\,K$, and is especially poor at $613\,K$.  In interpreting this disagreement, remember that we evaluated $\rho(T)$ in \cite{FLP} point by point from the viscosity data given in \cite{LU}, and then checked that these values were the same as those that the latter authors had used in scaling their strain rates.  Thus we did not use the nominal values of the temperature $T$ in any of those analyses.  Here, on the other hand, I have fit $\rho(T)$ by an analytic expression, Eqs.(\ref{rho-alpha}) and (\ref{CGalpha}), and have used this function of $T$ in plotting the curves shown in Fig.(5).  The problem is that the viscosity data for the lowest four temperatures shown in Fig.(2) does not fit onto a smooth curve.  Since there is no reason to believe that the material is undergoing any qualitative change in this temperature range \cite{JOHNSON}, we must presume that either the reported temperatures or the Newtonian viscosities -- or both -- are inaccurate.  Accordingly, instead of using the nominal temperatures  $593\,K$, $603\,K$, and $613\,K$ in drawing the theoretical curves in Fig.(5), I have used $588\,K$, $598\,K$, and $607\,K$ respectively, and have indicated these results by dashed lines.  Note that these small shifts in temperature produce large shifts in the predicted Newtonian viscosities, and thus move the data points closer to the smooth curve in Fig.(2).  More importantly, the low-temperature data is largely in the region where the linear Newtonian behavior is becoming nonlinear superplasticity; therefore the ability of the theory to account for the data seems significant. In short, although the effective temperature theory systematically departs from the strong scaling behavior obtained in \cite{FLP}, it appears that these departures are within the present uncertainties in the experimental data. 

Our one remaining point of contact with the data of Lu et al. \cite{LU} is the transient response shown in their constant strain-rate experiments.  As in \cite{FLP}, we can use these experimental results to obtain separate estimates of the parameters $\epsilon_0$ and $\tau_0$ instead of just their ratio.  To compute the corresponding stress-strain curves, write the equation of motion for the total strain $\epsilon^{total}$ (including elastic strain) in the form
\begin{equation}
{\sigma_y\over E}\,\dot{\tilde s} =\dot\epsilon_{xx}^{total} - {\epsilon_0'\over \tau_0}\,e^{-1/\chi}\,{\cal C}(\tilde s)\,\Bigl({\cal T}(\tilde s)-m\Bigr),
\end{equation}
and solve this simultaneously with Eqs. (\ref{dotm3}) and (\ref{chidot}) for $\tilde s$, $m$, and $\chi$ at fixed $\dot\epsilon_{xx}^{total}$.  In preparation for plotting stress-strain curves, we can let $\epsilon_{xx}^{total}$ replace time as the independent variable, in which case $\epsilon_0'$ appears separately as well as in the combination $\epsilon_0/\tau_0$.  Thus, fitting the transient response yields separate estimates for $\epsilon_0'$ and $\tau_0$.

The parameter $\epsilon_0'$ must be a large number in this version of the STZ theory, because the fraction of the volume covered by STZ's is proportional to $\epsilon_0'\,\exp\,(-1/\chi)$, not just to $\epsilon_0'$ by itself as in \cite{FLP}.  At $\chi=\chi_{\infty}$, this fraction would be unity if $\epsilon_0'$ were about $3\times 10^{14}$.  I find that $\epsilon_0'\cong 10^{14}$ works well for making the theoretical stress-strain curves agree with the experimental ones shown in \cite{LU}, Figs.(1) and (2).  With that value, the equilibrated fractional density, $\epsilon_0'\,\exp\,(-T_Z/T)$ is of order $.002$ at $T=648\,K$; thus the effective temperature theory produces estimates of the STZ density that are in accord with the idea that this density should be small, which was not necessarily the case in \cite{FLP}. With $\epsilon_0'\cong 10^{14}$, and our earlier estimate $\epsilon_0'/\tau_0 \cong 6.3\times 10^{19}\,{\rm sec}^{-1}$, we have $\tau_0 \sim 10^{-6}\,{\rm sec}$, which seems reasonable if we remember that $\tau_0^{-1}$ is the STZ transformation rate in the limit of infinite applied stress.  

With these parameters, plus $\kappa = 2$, $\beta=1$ and $c_0=1$ in Eq.(\ref{chidot}), and $E/\sigma_y \cong 50$, the stress-strain curves are essentially identical to those shown in Figs. (1) and (2) of \cite{FLP} which, in turn, were similar to the experimental ones shown in \cite{LU}.  In this fully nonlinear theory, the initial rise of the stress is accurately determined by Young's modulus, instead of being too small because the plastic response was unrealistically enhanced at small stresses in the quasilinear version.  Typical stress-strain curves, analogous to those in \cite{FLP}, are shown in Figs.(6) and (7).  As in \cite{FLP}, initial values of $\tilde s$ and $m$ are zero. I have assumed that the initial value of $\chi$ is always equal to $T/T_Z$, that is, that the samples are completely equilibrated initially by annealing at the experimental temperatures. 

Turn now to the thermodynamic properties of this theory, which were missing in the earlier version \cite{FLP} but now can be explored in detail.  Given our assumption that the configurational energy is simply proportional to the effective temperature, we can convert the equation of motion for $\chi$ in the absence of driving, i.e. Eq.(\ref{chidot}) with $\Gamma=0$, into an equation for the specific heat measured in a DSC experiment.  Let the heating rate be $h=\dot T$. Then (\ref{chidot}) becomes
\begin{equation}
\label{dchidT}
{d\chi\over dT}={\epsilon_0'\over c_0\,\tau_0}\,{\kappa\,\rho(T)\over h}\,e^{-\beta/\chi}\,\left({T\over T_Z}-\chi\right).
\end{equation}
The left-hand side is equal to the specific heat in units $C_D\,T_Z$.  Figure (8) shows $d\chi/dT$ computed by solving Eq.(\ref{dchidT}) with $\kappa = 2$, $\beta=1$ and $c_0=1$, and a heating rate of $10\,K$ per minute. The initial temperature used for integrating this equation was $550\,K$; but the results are insensitive to this value so long as it is low enough.  The different states of the system are specified by the initial values of $\chi$ which, in temperature units (i.e. expressed as $T_{eff}=T_Z\,\chi$) were chosen here to be $630\,K$, $640\,K$, $650\,K$, and $660\,K$. These curves resemble those found, for example, in De Hey et al. \cite{DEHEY} or Tuinstra et al. \cite{TUINSTRA} Eventually they should be compared with data for ${\rm Zr_{41.2}\,Ti_{13.8}\,Cu_{12.5}\,Ni_{10}\,Be_{22.5}}$ such as that shown in Fig. (2) of Busch et al. \cite{BUSCH}; but that analysis would best be carried out in connection with experiments like those discussed in the next paragraph.  Not shown in Fig.(8) is the prediction from Eq.(\ref{dchidT}) that, when the system is fully annealed at lower temperatures so that the initial $T_{eff}$ is well below $600\,K$, the spike becomes very sharp and moves to temperatures above $700\,K$.  

The differences between the areas under specific heat curves of the kind shown in Fig.(8) are equal to the differences between the enthalpies of systems with the corresponding  initial values of $\chi$.  Those values can be controlled experimentally by shearing the system at fixed bath temperatures and strain rates for long enough times that they achieve steady state.  On the theoretical side, we can compute the values of $\chi$ as functions of temperature and strain rate by finding the steady-state solutions of Eqs. (\ref{dotm3}) and (\ref{chidot}), as we have done to obtain the steady-state stresses in Figs. (4) and (5).  Thus our steady-state values of $\chi$ can be determined experimentally. Precisely such measurements have been performed by De Hey et al. \cite{DEHEY}, who interpreted their function $\chi$ as a reduced free volume instead of a reduced effective temperature.  The two interpretations may be effectively equivalent for systems held at constant pressure because, under that condition, the change in volume will be proportional to the effective temperature.      

Figure (4) in De Hey et al. \cite{DEHEY} shows values of $\chi$ as functions of strain rate at three different temperatures for thin ribbons of amorphous ${\rm Pd_{40}\,Ni_{40}\,P_{20}}$.  Without detailed information about other parameters of the kind obtained here from \cite{LU}, we cannot try to reproduce the results of \cite{DEHEY} theoretically.  Instead, I have used the parameters determined here for amorphous bulk ${\rm Zr_{41.2}\,Ti_{13.8}\,Cu_{12.5}\,Ni_{10}\,Be_{22.5}}$ to compute a graph analogous to the one in \cite{DEHEY}.  The results are shown in Fig.(9) for four different bath temperatures $T$.  Note that $\chi$ approaches $T/T_Z$ at low strain rates, and goes to $\chi_{\infty} = 0.032$ at large strain rates.  At intermediate rates, such as those shown in \cite{DEHEY}, the values of $\chi$ decrease as $T$ increases, consistent with the idea that the number of STZ's needed to sustain a fixed strain rate decreases when thermal fluctuations assist the transitions.  As anticipated in \cite{FLP}, these curves cross each other as they move to small strain rates.  Figure (9) implies that this crossover might be observed experimentally in ${\rm Zr_{41.2}\,Ti_{13.8}\,Cu_{12.5}\,Ni_{10}\,Be_{22.5}}$. 

All of the preceding calculations have been based on the equation of motion for $\chi$, Eq.(\ref{chidot}),  with the equilibration parameter $\beta$ set equal to unity.  Remember that the quantity $\beta\,k_B\,T_Z$ is a characteristic formation energy for configurational fluctuations that drive the effective temperature $T_{eff}$ toward the bath temperature $T$.  A value of $\beta$ smaller than unity implies that these fluctuations occur more frequently than the STZ's, which seems plausible.  (The opposite situation, $\beta > 1$, might also occur.) Figure (10) shows what happens to the steady-state stress versus strain rate curves if we choose $\beta = 0.5$.  In order to be at least roughly consistent with experimental data, that is, in order that the two terms on the right-hand side of Eq.(\ref{chidot}) be of comparable size when $\chi$ is near $\chi_{\infty}$, we must choose a much smaller value of $\kappa$ than previously. Specifically, $\kappa = 10^{-8}$ for the graphs shown in Fig.(10).  The most important new feature is that the curve for the lowest of the four temperatures shown here, $T= 573\,K$ no longer remains below the others as it does in Fig.(4), but now rises above and goes through a maximum and then a minimum before returning to approximately its previous behavior.  This multi-valued property is seen more clearly if, instead, we plot the reduced effective temperature $\chi$ as a function of the stress, as shown in Fig. (11). 
 
The multi-valued behavior of $\chi$ implies a shear-banding instability.  (See, for example, the analyses by Olmsted et al. \cite{OLMSTED1,OLMSTED2,OLMSTED3} of shear banding in several similar situations.) In the usual simple-shear experiment in a strip geometry, the shear stress remains constant across the sample in order to satisfy force-balance.  If the externally imposed shear rate is chosen so that the stress lies in the multi-valued region, then the sample will have to break up into regions of large and small flow or, equivalently, high and low effective temperature.  That is, the system will encounter a shear-banding instability and most likely will fail via shear fracture.  Figure (11) indicates that this instability appears only at temperatures lower than about $648\,K$, and that the onset stress increases with decreasing temperature.  The figure also implies that, even at low temperatures, uniform flow should be stable at sufficiently large driving forces. 

A satisfactory theory of shear banding also needs a length scale, because it must describe a smooth transition between the jammed and flowing regions of the material. The effective temperature theory suggests that a natural way to introduce this length is to add a diffusion term proportional to $\nabla^2\,\chi$ to the right hand side of Eq.(\ref{chidot}). The associated diffusion constant will be very much smaller than the ordinary thermal diffusion constant because configurational disorder must diffuse extremely slowly at temperatures below the glass transition. We may even be able to estimate the magnitude of this diffusion constant from the arguments presented in Sec. III.  Thus, the effective temperature theory seems to be giving us a clue about how to solve the long-standing problem of identifying an intrinsic length scale for shear localization. A fully detailed development of these ideas, however, is beyond the scope of the present paper.

\section{Concluding Remarks}  

My main conclusion is that the effective-temperature version of STZ theory looks promising but is far from being quantitatively confirmed by comparison with experimental data.  I see  several directions for future investigations.  

First, there is a need to combine mechanical and calorimetric measurements, in the manner described by De Hey et al.\cite{DEHEY}, in order to test predictions of the kind shown in Fig.(9).  Such experiments may come as close as is possible to actually measuring the effective temperature and learning whether it behaves as predicted.  It also would be useful to repeat the mechanical experiments with enough precision to test the predicted deviations from scaling shown in Figs.(4) or (10).  For this purpose, it might be well to use other materials such as polymeric glasses or, perhaps, colloidal suspensions in order to control the experimental conditions more precisely than seems possible with amorphous metals.

A second direction for further research is to develop the theory of shear banding along the lines described above and to test the results experimentally. For example, it should be possible to predict and measure the onset of spatial instability as a function of temperature and applied stress.

In my opinion, the principal theoretical question left unanswered is the temperature dependence of the rate factor $\rho(T)$.  This factor has been determined empirically here from experimental measurements of the Newtonian viscosity, with no theoretical justification whatsoever. The STZ theory described in this paper and in \cite{FLP} differs most markedly from the earlier flow-defect theories in that we ascribe the non-Arrhenius behavior of the viscosity, not to the density of defects, but to the transition rates.  Thus the kind of analysis that was used in deriving the Cohen-Grest formula \cite{COHEN-GREST} seems unlikely to be applicable.  Instead, we must learn how to perform a truly nonequilibrium analysis of the processes by which configurational rearrangements occur, both spontaneously and as driven by imposed stresses.  This is a large challenge -- perhaps the same as that of understanding the fundamental nature of the glass transition itself.

\begin{acknowledgments}
This research was supported primarily by U.S. Department of Energy Grant No. DE-FG03-99ER45762, and in part by the MRSEC Program of the National Science Foundation under Award No. DMR96-32716. I would like to thank W. Johnson for information about the data in Ref.\cite{LU}, and A. Liu, D. Durian, A. Lemaitre, L. Pechenik, A. Foglia, C. Maloney, and L. Mahadevan for useful discussions.  
\end{acknowledgments}

\begin{figure}
      \includegraphics[angle=0, width=0.7\columnwidth]{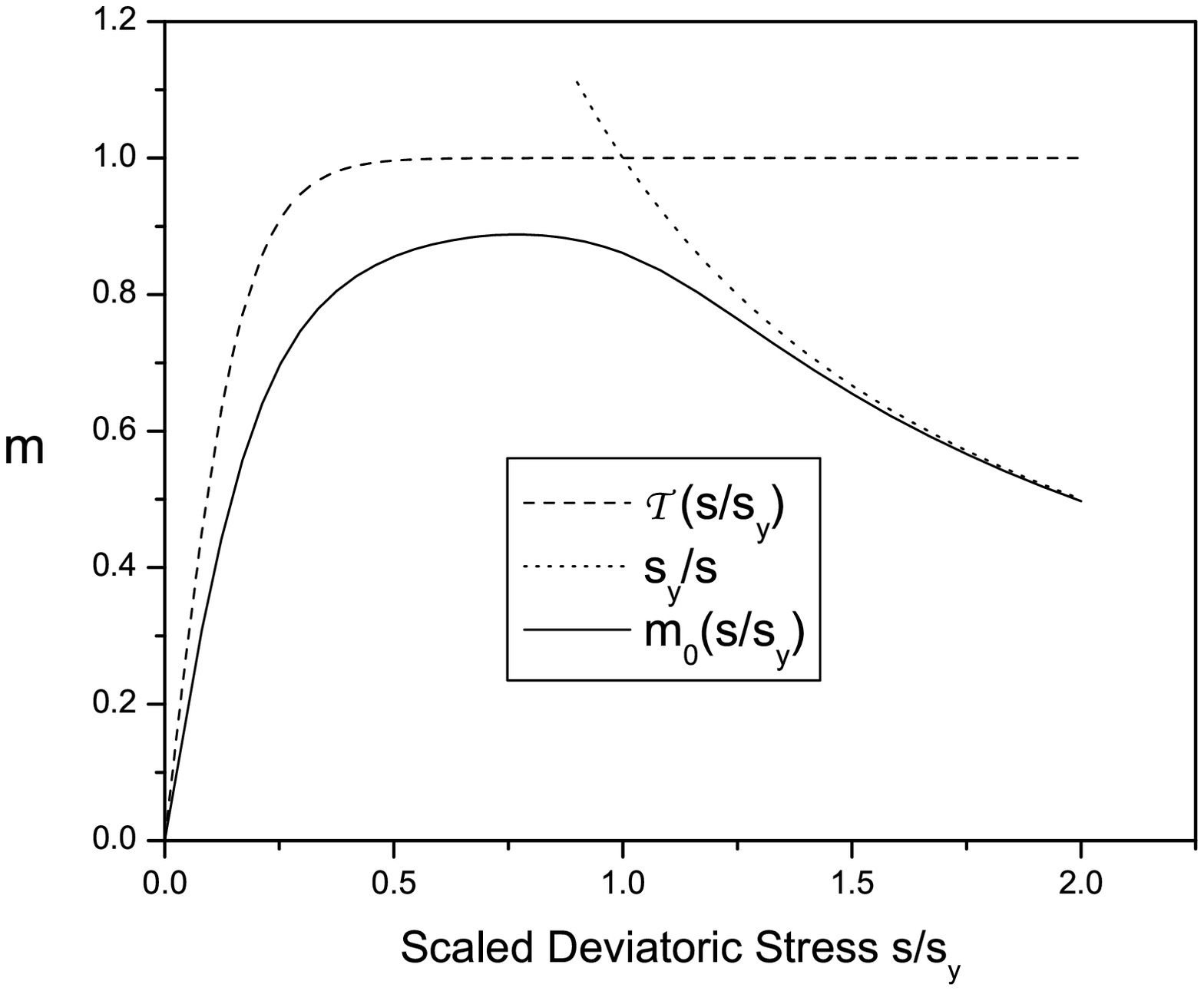} 
\caption{Graph of the function $m_0(\tilde{s})$, $\tilde{s} = s/\bar{\mu} 
\sim s/s_y$, for $\alpha = 6$.  
Also shown are the related functions ${\cal T}(\tilde{s})$ and  $1/\tilde{s}$. }
\end{figure}

\begin{figure}
      \includegraphics[angle=0,
      width=0.7\columnwidth]{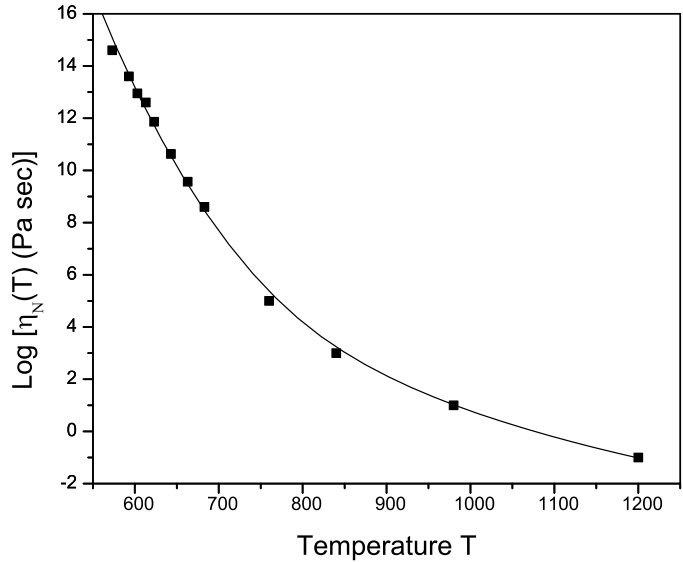} 
\caption{Experimental values of the Newtonian viscosity $\eta_N(T)$ taken from Lu et al.,\cite{LU} and the analytic fit to these points obtained by choosing parameters in Eqs.(\ref{rho-alpha}) and (\ref{CGalpha}) as explained in the text.}
\end{figure}

\begin{figure}
      \includegraphics[angle=0,
      width=0.7\columnwidth]{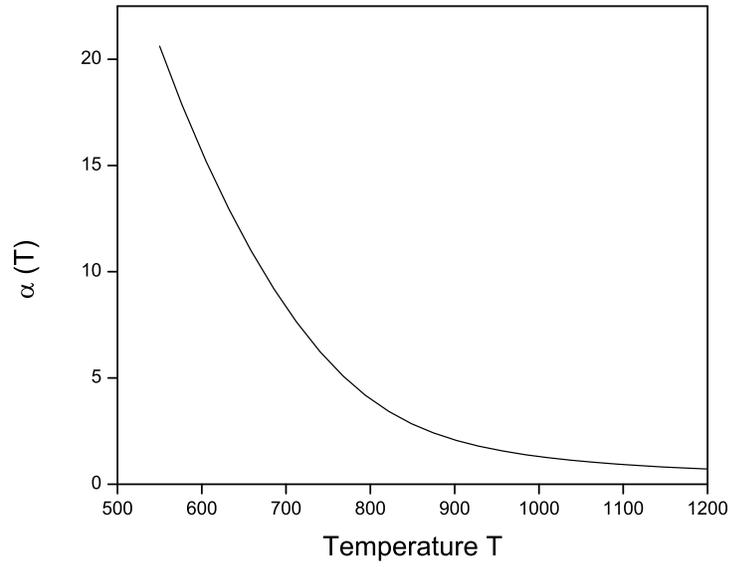} 
\caption{The function $\alpha(T)$ determined by fitting the viscosity data shown in Fig.(2). }
\end{figure}

\begin{figure}
      \includegraphics[angle=0,
      width=0.7\columnwidth]{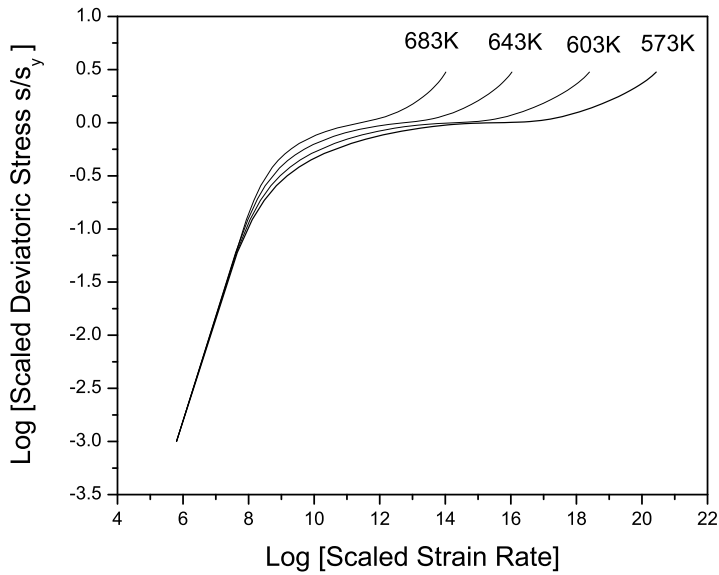} 
\caption{Graphs of the scaled deviatoric stress $\tilde s = s/s_y$ as functions of the scaled, steady-state strain rate $\eta_N(T)\,\dot\epsilon_{xx}^{pl}$ for four different temperatures.}
\end{figure}

\begin{figure}
      \includegraphics[angle=0,
      width=0.7\columnwidth]{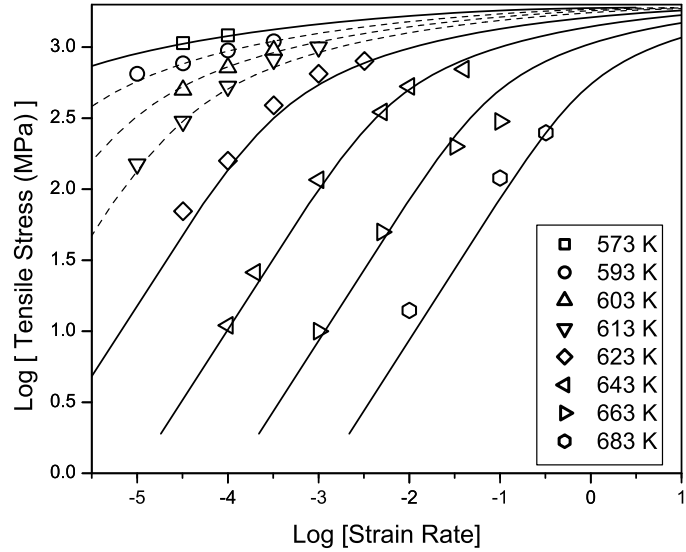} 
\caption{Tensile stress as a function of the steady-state strain rate $\dot\epsilon_{xx}^{pl}$ (in ${\rm sec}^{-1}$). The data points, taken from Lu et al.\cite{LU}, correspond to the nominal temperatures as shown, and the solid theoretical curves are computed using those temperatures.  The dashed curves are computed using, from left to right, $T= 588\,K$, $598\,K$, and $607\,K$.}
\end{figure}

\begin{figure}
      \includegraphics[angle=0,
      width=0.7\columnwidth]{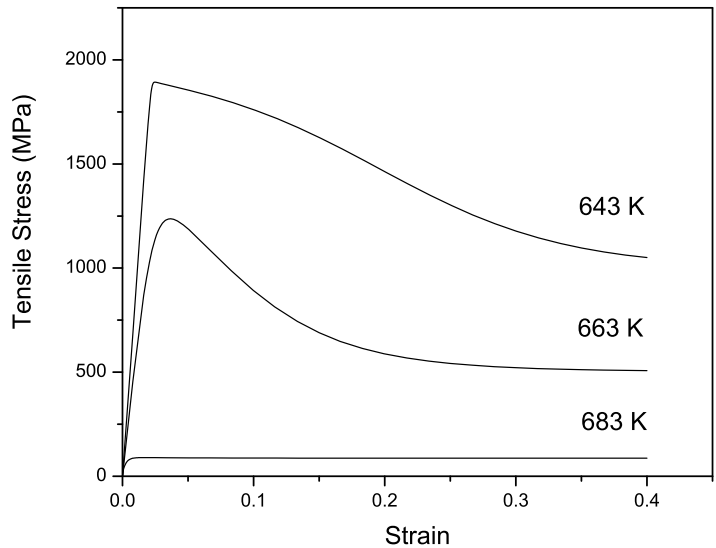} 
\caption{Theoretical stress-strain curves for $\dot\epsilon_{xx}^{total} = 0.1 \,\,{\rm sec}^{-1}$ at three different temperatures.}
\end{figure}

\begin{figure}
      \includegraphics[angle=0,
      width=0.7\columnwidth]{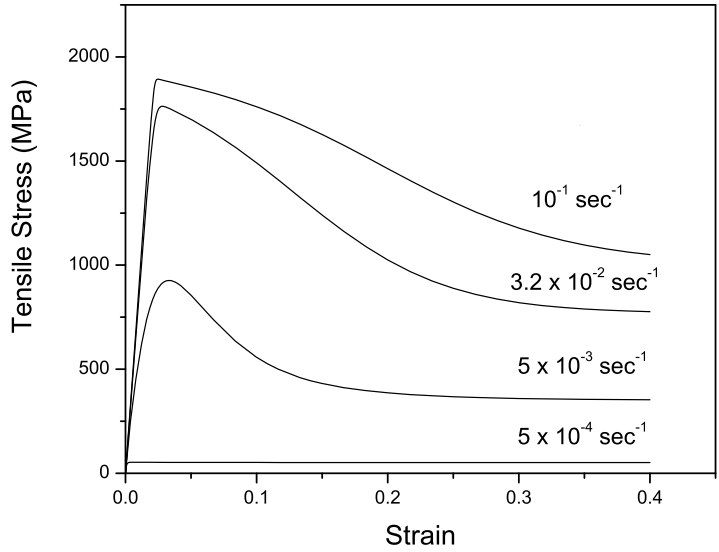} 
\caption{ Theoretical stress-strain curves for $T= 643\,K$ at four different strain rates $\dot\epsilon_{xx}^{total}$.}
\end{figure}

\begin{figure}
      \includegraphics[angle=0,
      width=0.7\columnwidth]{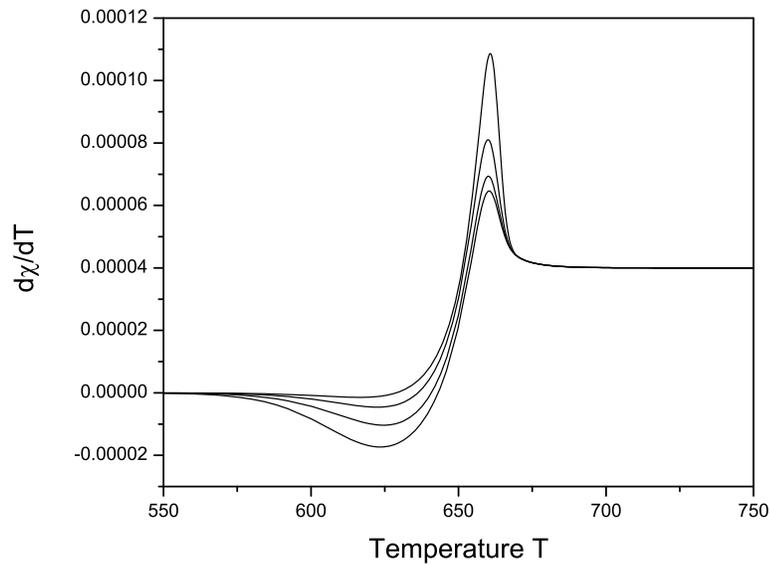} 
\caption{Scaled specific heat curves, $d\chi/dT$, corresponding to simulated DSC measurements at a heating rate of $10\,K$ per minute.  The initial states, in order of decreasing peak height, have effective disorder temperatures $T_{eff}= 630\,K$, $640\,K$, $650\,K$, and $660\,K$.}
\end{figure}

\begin{figure}
      \includegraphics[angle=0,
      width=0.7\columnwidth]{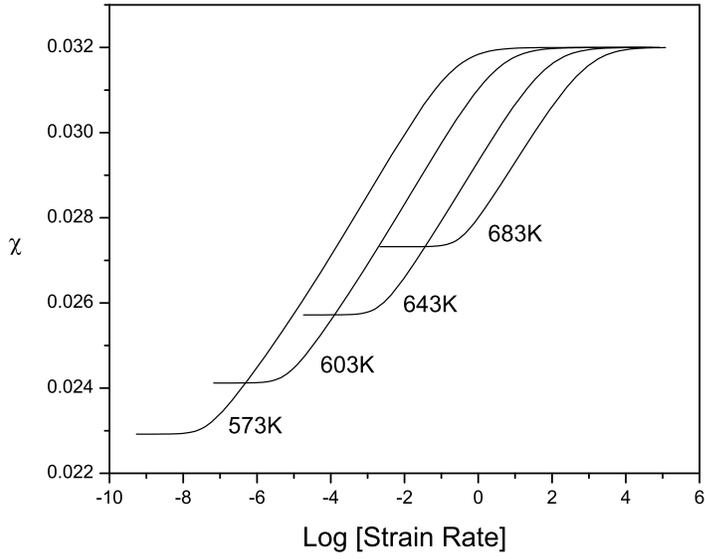} 
\caption{Steady-state values of the scaled effective temperature $\chi$ as functions of strain rate at four different temperatures.}
\end{figure}

\begin{figure}
      \includegraphics[angle=0,
      width=0.7\columnwidth]{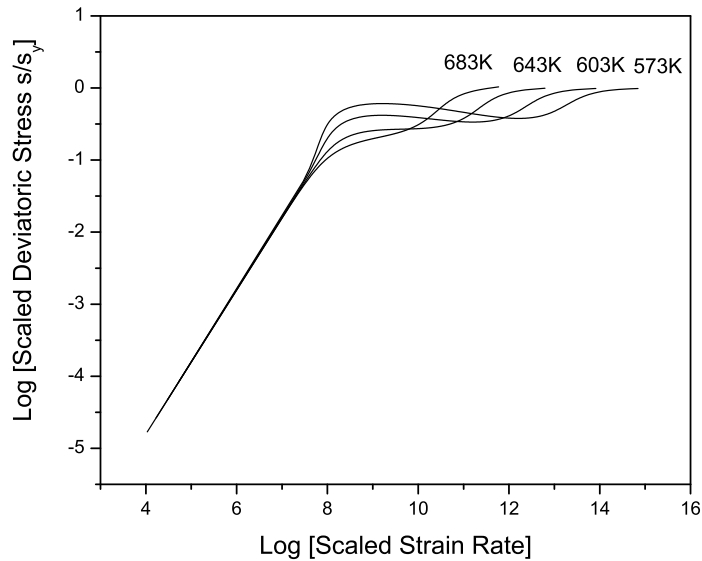} 
\caption{Graphs of the scaled deviatoric stress $\tilde s = s/s_y$ as functions of the scaled strain rate $\eta_N(T)\,\dot\epsilon_{xx}^{pl}$ for four different temperatures. This figure is analogous to Fig.(4), but the equilibration parameter $\beta$ has been set equal to $0.5$.}
\end{figure}

\begin{figure}
      \includegraphics[angle=0,
      width=0.7\columnwidth]{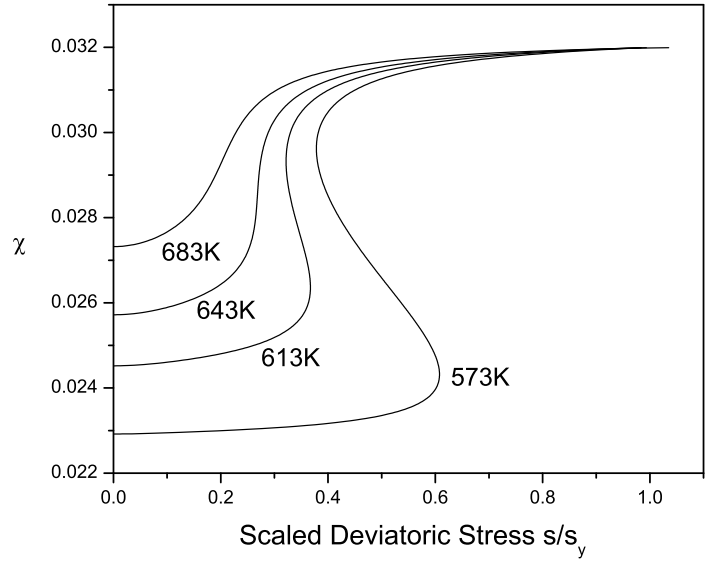} 
\caption{Graphs of the scaled effective temperature $\chi$ as functions of the scaled deviatoric stress $\tilde s = s/s_y$ at four different temperatures. As in Fig.(10), the equilibration parameter is $\beta = 0.5$.}
\end{figure}

\end{document}